\documentclass[prd, twocolumn, nofootinbib, floatfix]{revtex4-1}
\usepackage[mathscr]{euscript}
\usepackage{amsmath}
\usepackage{graphicx}
\usepackage{dcolumn}
\usepackage{bm}
\usepackage{epsfig}
\usepackage{amssymb,latexsym,mathrsfs}
\usepackage{graphicx}
\usepackage{color}
\usepackage{hyperref}
\usepackage{float}
\usepackage{diagbox}

\hypersetup{
    colorlinks=true,
    linkcolor=red,
    citecolor=blue,
}

\usepackage{amsmath}
\usepackage{amssymb}
\usepackage{subfigure}
\usepackage{hyperref}
\usepackage{url}
\usepackage{xcolor}
\usepackage{color}
\definecolor{amaranth}{rgb}{0.9, 0.17, 0.31}
\definecolor{purple(munsell)}{rgb}{0.62, 0.0, 0.77}
\definecolor{americanrose}{rgb}{1.0, 0.01, 0.24}
\definecolor{palatinateblue}{rgb}{0.15, 0.23, 0.89}
\definecolor{royalblue(web)}{rgb}{0.25, 0.41, 0.88}
\definecolor{hanpurple}{rgb}{0.32, 0.09, 0.98}
\definecolor{beaublue}{rgb}{0.74, 0.83, 0.9}
\definecolor{carminered}{rgb}{1.0, 0.0, 0.22}
\definecolor{brightpink}{rgb}{1.0, 0.0, 0.5}
\definecolor{vividviolet}{rgb}{0.62, 0.0, 1.0}
\hypersetup{ linktoc=all,
    colorlinks, linkcolor={palatinateblue},
    citecolor={brightpink}, urlcolor={amaranth}}

\newcommand{\be}{\begin{equation}}
\newcommand{\ee}{\end{equation}}
\newcommand{\bs}{\begin{split}} 
\newcommand{\bea}{\begin{eqnarray}}
\newcommand{\eea}{\end{eqnarray}}

\newcommand{\bes}{\begin{subequations}}
\newcommand{\ees}{\end{subequations}}

\renewcommand{\d}[1]{\ensuremath{\operatorname{d}\!{#1}}}

\renewcommand{\d}[1]{\ensuremath{\operatorname{d}\!{#1}}}

\newcommand{\bo}{\raise-1mm\hbox{\Large$\Box$}}

\newcommand{\bd}{\boldsymbol}

\begin{document}

\title{Quantum power distribution of relativistic acceleration radiation: classical electrodynamic analogies with perfectly reflecting moving mirrors}
\author{Abay Zhakenuly${}^{1}$}
\author{Maksat Temirkhan${}^{1}$}
\author{Michael R.R. Good${}^{1}$}
\author{Pisin Chen${}^{2,3}$}
\affiliation{${}^{1}$Physics Department \& Energetic Cosmos Laboratory, Nazarbaev University, Nur-Sultan, 010000, Qazaqstan.}
\affiliation{${}^{2}$ Department of Physics and Center for Theoretical Sciences, National Taiwan University, Taipei, 10617, Taiwan}
\affiliation{${}^{3}$LeCosPA, National Taiwan University, Taipei, 10617, Taiwan}

\begin{abstract} 
We find the quantum power emitted and distribution in $3+1$-dimensions of relativistic acceleration radiation using a single perfectly reflecting mirror via Lorentz invariance demonstrating close analogies to point charge radiation in classical electrodynamics.    
\end{abstract} 

\date{\today} 

\maketitle



\section{Introduction}

There are deep connections between point charge radiation in classical electrodynamics and quantum vacuum acceleration radiation from perfectly reflecting point mirrors of DeWitt-Davies-Fulling \cite{DeWitt:1975ys,Davies:1976hi,Davies:1977yv}. For instance in 1982 Ford-Vilenkin \cite{Ford:1982ct} demonstrated that the force on a $1+1$ moving mirror has the same covariant expression as the Lorentz-Abraham-Dirac (LAD) radiation reaction force of an arbitrary moving point charge in $3+1$ classical electrodynamics. The connection extends to scalar source changes demonstrated, for instance, in 1992 by Higuchi-Matsas-Sudarsky \cite{Higuchi:1992td} through the discovery that photon emission from a uniformly accelerated classical charge in the Minkowski vacuum corresponds to emission of a zero-energy Rindler photon into the Unruh thermal bath. In 1994 Hai \cite{hai} found  the quantum energy flux integrated along a large sphere in the asymptotic future as the Larmor formula for the power radiated by a moving scalar charge with respect to an inertial observer.  In 2002 Ritus \cite{Ritus:2003wu,Ritus:2002rq,Ritus:1999eu,Nikishov:1995qs} found symmetries linking creation of pairs of massless bosons or fermions by an accelerated mirror in 1+1 space and the emission of single photons or scalar quanta by electric or scalar charges in 3+1 space.  

The above studies are just some of the fascinating connections, nowhere near exhaustive, so far discovered.
Recent studies (\cite{doi:10.1098/rspa.2020.0656,Landulfo:2019tqj,1819893}) have also strengthened the connection between quantum acceleration radiation and classical radiation of point charges, namely the works by Landulfo-Fulling-Matsas who found \cite{Landulfo:2019tqj} zero-energy Rindler modes are not mathematical artifacts but
are critical to understanding the radiation in both the classical and quantum realm, confirming that Larmor radiation emitted by a charge can be seen as a consequence of the Unruh thermal bath.  Moreover, Cozzella-Fulling-Landulfo-Matsas \cite{1819893} conclude that uniformly accelerated pointlike structureless sources emit only zero-energy Rindler particles using Unruh-deWitt detectors \cite{PhysRevD.14.870}. The analogies hold true with respect to the uniformly accelerated moving mirror which emits zero energy flux but produce non-zero particle counts as computed from the beta Bogolubov coefficients (e.g. \cite{Birrell:1982ix}). 

Investigations are underway aimed at extending the $1+1$ dimensional moving mirror model to $3+1$ dimensions, and understanding the production of scalar particles in the relativistic regime \cite{Lin:2020itp}.  These efforts are carried out with the goal of direct detection of relativistic moving mirror radiation \cite{Chen:2015bcg,Chen:2020sir,Brown2015}, which complement the growing accumulation of observations of the dynamical Casimir effect (see references therein \cite{Dodonov:2020eto}).

Experimental verification will be facilitated by knowing the distribution of detected radiation.  Recent studies have only just solved for the five-classes of uniformly accelerated trajectory distribution \cite{Good:2019aqd} in classical electrodynamics (effective Unruh-like temperatures have also been calculated \cite{Good:2020hav}).  Unfortunately, mathematically, no settled upon or convincing \cite{kirk} covariant expression has yet been derived to express the power distribution in a frame-independent formulation \cite{Rohrlich:1097602}. In order to know the distribution of quantum power from a relativistic moving mirror we are also forced to abandon cherished covariant language, while at the same time, maintaining the principle of Lorentz invariance.   

In this work we compute the quantum power Larmor formula for relativistic moving mirror radiation, the explicit non-covariant quantum power distribution, and apply the results to the simple case of abrupt mirror creation, violent acceleration, and near-instantaneous final constant velocity state of motion for the spectrum of radiation of the mirror. Along the way, we highlight the direct analogies to classical radiation from a moving point charge in electrodynamics.

Our paper is organized as follows: in Sec.\ \ref{sec:power}, we obtain a total power definition for quantum radiation emitted by a single relativistic moving mirror in $1+1$-dimensions. We highlight that it has identical form as the relativistic generalization of the  Larmor formula for power emitted by an accelerated point charge.   In Sec.\ \ref{sec:angtime}, we derive the angular distribution in time of the quantum radiation of the mirror in $3+1$ dimensions using Lorentz invariance.  Here we utilize the ansatz that proper acceleration magnitude is a Lorentz scalar independent of direction or  dimension. We determine the quantum power distribution has the same form as in classical electrodynamics. In Sec.\ \ref{sec:angfreq}, we derive the radiation integral demonstrating the approach in electrodynamics equally applies to the mirror. Finally, in Sec.\ \ref{sec:betadecay}, we specialize our results to an abruptly created, rapidly accelerated, constant velocity moving mirror connecting the frequency independent spectrum with that of beta decay.  We discuss and conclude in Secs. \ref{sec:dis} and \ref{sec:conc}. Throughout we use natural units, $\hbar = c = 1$.
For conversion from the SI electrodynamics analog, one requires: $q \to 1$, $\mu_0 \to 1$, $\epsilon_0 \to 1$. For the commonly used Gaussian units one converts by $4\pi \epsilon_0 \to 1$ and $\mu_0 \to 4\pi$.  \\

\section{Relativistic quantum Larmor formula}\label{sec:power}
The energy flux, $\mathcal{F}$, radiated by the mirror moving along a trajectory $p(u)$ in null coordinates where $u=t-x$ and $p$ is the advanced time, $v=t+x$, position, is derived from the Davies-Fulling quantum stress tensor \cite{Davies:1976hi},
\be \mathcal{F}^R(u) = -\frac{1}{24\pi} \{p(u),u\},\ee
where the total energy emitted to the right of the mirror is (e.g. Walker \cite{walker1985particle})
\be E^R = \int_{-\infty}^{\infty} \mathcal{F}^R \d u,\label{rightE}\ee
defining the Schwarzian derivative by (e.g. Fabbri-Navarro-Salas \cite{Fabbri})
\begin{align}
\big\{p(u), u \big\} &= \frac{p'''}{p'} - \frac{3}{2} \left( \frac{p''}{p'} \right)^2.
\end{align}
This energy Eq.~(\ref{rightE}) can be expressed in lab coordinates as (see Eq. 2.34 of Good-Anderson-Evans \cite{good2013time}),
\be E^R = \frac{1}{12\pi} \int_{-\infty}^{\infty} \alpha^2(1+\dot{x})\,\d t,\label{parts1}\ee
where $\alpha$ is the proper acceleration of the mirror.  The energy radiated to the left, $E^L$, is found by the same expression but with a parity flip, $\dot{x} \rightarrow -\dot{x}$, so that
\be E^L = \frac{1}{12\pi} \int_{-\infty}^{\infty} \alpha^2(1-\dot{x})\,\d t,\ee
which gives a total radiated energy
\be E = E^R+E^L =  \frac{1}{6\pi} \int_{-\infty}^{\infty} \alpha^2\,\d t.\label{ET}\ee
This allows us to identify and define a relativistic quantum power for the moving mirror, $P \equiv \d E/\d t$,
\be E = \int_{-\infty}^{\infty} \frac{d E}{d t}\d t \equiv \int_{-\infty}^{\infty} P \d t,\ee
which gives, from Eq.~(\ref{ET}), a familiar acceleration scaling:
\be P = \frac{\alpha^2}{6\pi}.\label{P}\ee
The quantum vacuum scalar radiation power, Eq.~(\ref{P}), (in SI units $P_{\textrm{mirror}} = \hbar \alpha^2/6\pi c^2$) emitted by the mirror enjoys the same scaling as the relativistic Larmor formula,
\be P_{\textrm{electron}}= \frac{2}{3}\frac{q^2\alpha^2}{4\pi \epsilon_0 c^3}=\frac{q^2\alpha^2}{6\pi \epsilon_0 c^3} \to \frac{q^2\alpha^2}{6\pi},\ee
in classical electrodynamics, where we start in SI units and ``$\to$" implies conversion to natural units where $\epsilon_0 =c=1$, and $\alpha$ is the magnitude of the proper acceleration of the moving point charge (e.g. electron).

\begin{figure}[ht]
\begin{center}
{\rotatebox{0}{\includegraphics[width=2.6in]{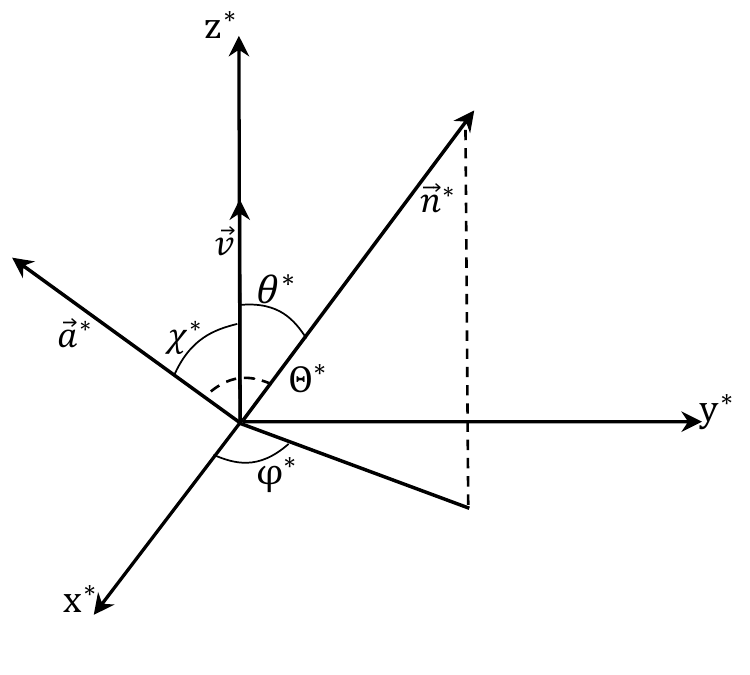}} {\caption{\label{fig1} This figure shows the angles we adopt between the relevant vectors, where $\chi^*$ is the angle between acceleration and velocity in the $x^*z^*$ plane, $\theta^*$ is the angle between velocity and the normal vector, $\Theta^*$ is the angle between acceleration and the normal vector, and $\phi^*$ is the azimuthal angle. }}} 
\end{center}
\end{figure}

\section{Angular distribution in time}\label{sec:angtime}
The relativistic quantum generalization of Larmor's power formula, Eq.~(\ref{P}), is a Lorentz invariant scalar, 
\be P = P^* = \frac{\alpha^2}{6\pi},\label{P1}\ee
where $P^*$ is the instant rest frame power and $P$ is the lab frame power.  Here we make a critical assumption about the $1+1$-dimensional result, Eq.~(\ref{P1}): the universality of the Lorentz invariant scalar in any frame suggests it is independent of dimension.  It is just a number after all, with no associated direction (see Appendix \ref{LIscalar} for elicitation). We take this as an ansatz and find it possible to proceed.  In which case, a simple definition for the angular distribution of the power Eq.~(\ref{P1}), of a moving mirror in $3+1$ dimensions, can be written
\be P = \int \frac{dP^*}{d\Omega^*}d\Omega^* = \int \frac{dP^*}{d\Omega^*}\sin\Theta^*d\Theta^*d\phi^*.\label{ansatz}\ee
Here the $*$ designates the instantaneous rest frame, the angle $\Theta^*$ is between the acceleration 3-vector $\bd{a}^*$ and the direction to the observer in the $*$ frame. What we have done is assumed that the power formula derived in $1+1$ dimensions, Eq.~(\ref{P1}), holds true in $3+1$ dimensions (this was known 20 years ago by Bekenstein for power emitted by black holes in the context of information transmission \cite{Bekenstein:2001tj}).  Likewise we will assume the power is distributed in $3+1$ dimensions by using the $3+1$ dimensional Lorentz covariant definition of proper acceleration.  Therefore \be \frac{dP^*}{d\Omega^*} = \frac{|\bd{a}^*\times \hat{\bd{n}}^*|^2}{16\pi^2} =  \frac{\alpha^2\sin^2\Theta^*}{16\pi^2},\label{dP*}\ee
where the instantaneous rest frame unit vector, $\hat{\bd{n}}^*$, is
\be \hat{\bd{n}}^* = \sin\theta^*\cos\phi^* \hat{x}^* + \sin\theta^*\sin\phi^* \hat{y}^* + \cos\theta^*\hat{z}^*,\ee
and $\bd{a}^*$ is the proper acceleration 3-vector in the instant rest frame, where $\alpha^2 \equiv -a_\mu a^\mu$. The acceleration 4-vector in the instant rest frame has components $(0,\bd{a}^*)$, i.e. $\alpha$ is the invariant Lorentz scalar; the acceleration felt by the moving mirror itself; its `property' \cite{Rindler:108404}. The numerical $16\pi^2$ factor in Eq.~(\ref{dP*}) originates from
\be \int_0^{2\pi}\int_0^\pi \frac{\sin^2\Theta^*}{16\pi^2}\sin\Theta^*d\Theta^*d\phi^* = \frac{1}{16\pi^2}\frac{8\pi}{3} = \frac{1}{6\pi}.\ee
See Figure \ref{fig1} for an illustration of the usual angle convention we have adopted.  We now wish to find the power distribution starting from Eq.~(\ref{dP*}). We will find the dimensional allocation without appeal to fields or potentials, using only the principle of Lorentz invariance.  

First, let us transform the solid angle to the lab frame. Under Lorentz transform the solid angle is (e.g. \cite{kirk})
\be d\Omega^* = \frac{d\Omega}{\gamma^2(1-\beta\cos\theta)^2},\ee
and the energy density is $dU^* = \gamma (1-\beta\cos\theta)dU$, while the square of the acceleration scalar is written,
\be \alpha^2 = \gamma^6(a^2 - (\bd{\beta}\times \bd{a})^2).\ee
Hence, we write, using $dt = \gamma dt^*$, the distribution from the source, Eq.~(\ref{dP*}), in the lab frame coordinates \be \frac{dP}{d\Omega} = \frac{1}{\gamma^4(1-\beta\cos\theta)^3}\frac{dP^*}{d\Omega^*},\ee
obtaining,
\be \frac{dP}{d\Omega} =\frac{1}{16\pi^2} \frac{\gamma^2[a^2-(\bd{\beta}\times \dot{\bd{\beta}})^2]}{(1-\beta\cos\theta)^3}\sin^2\Theta^*.\label{dPdO}\ee
Since the distribution is needed in the lab frame, the rest  frame angle $\Theta^*$  should be described in terms of  lab frame angles $\theta$, $\phi$ and $\chi$ (see Appendix \ref{appB}). To put Eq.~(\ref{dPdO}) in a more explicitly useful (and recognizable) form without reference to the angle $\Theta^*$, we would like to show that
\be g^{2}\gamma^2[a^2-(\bd{v}\times \bd{a})^2]\sin^2\Theta^* = [\hat{\bd{n}} \times ((\hat{\bd{n}}-\bd{v})\times \bd{a})]^2, \ee
or more concisely, that  
\be |\bd{a}^*\times \hat{\bd{n}}^*|^2= \frac{\gamma^4}{g^2}[\hat{\bd{n}} \times ((\hat{\bd{n}}-\bd{v})\times \bd{a})]^2.\label{crossto2}\ee
To do this, we express the instantaneous rest frame 3-vector acceleration, $\bd{a}^*$, in terms of the lab frame 3-vector acceleration, $\bd{a}$, via the appropriate usual Lorentz acceleration transformation (e.g. \cite{Rahaman:1968858})
\be \bd{a}^* = \frac{\bd{a}}{\gamma^2(1-v^2)^2} - \frac{(\bd{a}\cdot \bd{v})\bd{v}(\gamma-1)}{v^2\gamma^2(1-v^2)^3}+\frac{(\bd{a}\cdot \bd{v})\bd{v}}{\gamma^2(1-v^2)^3}.\ee
We can transform instant rest angles $(\theta^*,\phi^*)$ to lab angles $(\theta,\phi)$ by
\be \sin^2\theta^* = \frac{\sin^2\theta}{\gamma^2 g^2}, \quad \cos^2\theta^* = \frac{(\cos\theta-\beta)^2}{g^2},\ee
where $g$ is the Doppler factor, $g\equiv 1-\beta\cos\theta$ and $\phi^* = \phi$.  This gives,
\be |\bd{a}^*\times \hat{\bd{n}}^*|^2=|\hat{\bd{n}}(\hat{\bd{n}}\cdot \bd{a}) - \bd{a} - \bd{v}(\hat{\bd{n}}\cdot \bd{a}) +\bd{a}(\hat{\bd{n}}\cdot\bd{v})|^2\frac{\gamma^4}{g^2}.\label{a*n*}\ee
From $\bd{A}\times(\bd{B}\times\bd{C}) = \bd{B}(\bd{A}\cdot\bd{C})-\bd{C}(\bd{A}\cdot\bd{B})$, the right hand side of Eq.~(\ref{a*n*}) is the right hand side of Eq.~(\ref{crossto2}). Therefore our power distribution in the lab frame, Eq.~(\ref{dPdO}), is (see Appendix \ref{appC} for a demonstration of well-known limits)
\be \frac{dP}{d\Omega} =\frac{1}{16\pi^2} \frac{[\hat{\bd{n}} \times ((\hat{\bd{n}}-\bd{\beta})\times \dot{\bd{\beta}})]^2}{(1-\beta\cos\theta)^5}.\label{dpdO}\ee
This form for the quantum power distribution of scalar radiation from a relativistic moving mirror is that of the widely used textbook result for the classical power distribution of an arbitrarily moving electric point charge  (e.g. \cite{Jackson:490457,Zangwill:1507229,Griffiths:1492149}).  Notice this derivation does not rely upon fields, or potentials and depends solely on the validity of Lorentz invariance of proper acceleration and its vector decomposition in $3+1$-dimensions.  Although the derivation method of Eq.~(\ref{dpdO}) was for quantum radiation from a moving mirror, it is also valid for classical radiation from a moving electron. 
\section{Angular distribution in frequency}\label{sec:angfreq}
In this section we  confirm a procedure to obtain the widely-used classical radiation integral of electrodynamics but for the case of quantum scalar radiation from a moving mirror.  This is needed to find the angular distribution in frequency.  We demonstrate that the derivation and assumptions applied in the context of classical electromagnetic fields is applicable to the quantum scalar field.  

Converting to the frequency domain, requires considering the power distribution in the time domain as an energy density distribution in both time and angle,
 \be \frac{d P(t)}{d\Omega} = \frac{d^2 U}{dtd\Omega},\label{Pt}\ee
 where we write 
 \be \frac{d U}{d\Omega}= \int\displaylimits_{-\infty}^{+\infty} dt\, \frac{dP(t)}{d\Omega} = \int\displaylimits_0^\infty d\omega \frac{dI(\omega)}{d\Omega},\ee
defining the angular distribution in frequency,
 \be \frac{d I(\omega)}{d\Omega} = \frac{d^2 U}{d\omega d\Omega}.\label{I omega}\ee
 Parsevel's theorem can be used to deduce that 
 \be \frac{d U}{d \Omega}=\int\displaylimits_{-\infty}^{+\infty} dt\, \frac{dP(t)}{d\Omega}=\frac{1}{2 \pi} \int\displaylimits_{-\infty}^{+\infty} d\omega\, \frac{dP(\omega)}{d\Omega}. \ee
We apply a reality condition to the integrand such that $P(t) = P^*(t)$ and $P(\omega) =P^*(-\omega)$, which gives an even integral,
\be \frac{dU}{d\Omega} = \frac{1}{\pi} \int\displaylimits_{0}^{+\infty}d\omega\, \frac{dP(\omega)}{d\Omega}.\ee
The angular distribution in frequency can be found by a derivative, 
\be \frac{d}{d \omega} \frac{d U}{d \Omega}=\frac{d}{d \omega} \frac{1}{ \pi} \int d\omega\, \frac{dP(\omega)}{d\Omega}, \ee
which amounts to, using the notation of Eq.~(\ref{I omega}),
\be \frac{dI(\omega)}{d\Omega}= \frac{d^2 U}{d\omega d \Omega}= \frac{1}{ \pi}\, \frac{dP(\omega)}{d\Omega}. \label{thepi} \ee
We now have picked up a $\pi$ and a Fourier transform.  We plug in Eq.~(\ref{dpdO}) which is the time-domain Eq.~(\ref{Pt}), into 
\be \frac{dP(\omega)}{d\Omega} = \int\displaylimits_{-\infty}^{+\infty} dt\, \frac{dP(t)}{d\Omega} e^{i\phi},\ee
where $\phi = \omega (t_r - \hat{\bd{n}}\cdot \bd{r}_0(t_r))$, using the radiation zone approximation.\footnote{This assumes the observation point is very far from the regions of space where the acceleration is non-zero: $\hat{\bd{n}}_r \approx \hat{\bd{r}}$ is constant.} The mirror always moves on some arbitrary trajectory $\bd{r}_0(t)$ with velocity $\bd{v}(t) = \dot{\bd{r}_0}(t)$.   Then using
Eq.~(\ref{thepi}), as well as expressing the Fourier transform over retarded time, $dt = g dt_r$, gives  the widely-used analog formula, see e.g. Jackson \cite{Jackson:490457} or Zangwill \cite{Zangwill:1507229}, 
\be \frac{dI(\omega)}{d\Omega} = \frac{1}{16\pi^3} \left|\int\displaylimits_{-\infty}^{\infty}dt_r\frac{[\hat{\bd{n}} \times ((\hat{\bd{n}}-\bd{\beta})\times \dot{\bd{\beta}}]}{(1-\beta\cos\theta)^2} e^{i\phi}\right|^2.\label{Iw}\ee
An important virtue of the form of Eq.~(\ref{Iw}) is that the integrand is zero when the mirror acceleration is zero, which will always be the case for collision scattering-type and open-orbit situations where the mirror is subjected to a force for a finite amount of time.  In the next section we will look specifically at such a situation in the case of abrupt creation of the mirror itself.

\section{Specialized Case: abrupt mirror creation} \label{sec:betadecay}
The sudden creation of a rapidly moving mirror will be accompanied by the emission of radiation.  The mirror will be initially at rest and violently accelerated during a short time interval to its final velocity. We wish to calculate the spectrum of this perfectly reflecting accelerated boundary with violent acceleration to constant velocity. Previously, Brown and Louko described the case of smooth creation of a two-sided Dirichlet mirror in (1+1)-dimensional flat space-time that generates flux of real quanta \cite{Brown2015}. 

To compute the spectrum we first start with the use of a perfect differential identity (see Appendix \ref{appA} for a proof),
\be \frac{\hat{\bd{n}} \times ((\hat{\bd{n}}-\bd{\beta})\times \dot{\bd{\beta}}}{(1-\beta\cos\theta)^2} = \frac{d}{dt_r} \left[ \frac{\hat{\bd{n}}\times(\hat{\bd{n}}\times \bd{\beta})}{1-\beta\cos\theta} \right],\label{totald}\ee
where the derivatives are evaluated at retarded time. We apply this identity to perform an integration by parts on Eq.~(\ref{Iw}),
\bea 
\frac{dI(\omega)}{d\Omega} &=& \frac{1}{16\pi^3}\left|\left.\frac{\hat{\bd{n}}\times(\hat{\bd{n}}\times \bd{\beta})}{1-\beta\cos\theta} e^{i\phi}\right|_{0}^{+\infty} \right.\\
&&\quad-\left. i\omega \int\displaylimits_{0}^{\infty} d t_r\, \hat{\bd{n}}\times(\hat{\bd{n}}\times \bd{\beta})e^{i\phi}\right|^2,\label{parts}
\eea
where the boundary terms vanish.   Again, we have defined $\phi =\omega t_r -\bd{k}\cdot\bd{r}_0(t_r)$, where $k=\omega$. Using $\hat{\bd{n}} \approx \hat{\bd{r}}$ as a constant vector means it comes outside the integral, and since $|\hat{\bd{r}}\times(\hat{\bd{r}}\times \bd{\beta})|^2 = |\hat{\bd{r}}\times \bd{\beta}|^2$ for any vector $\bd{\beta}$, we obtain the angular spectrum of radiated energy as
\be \frac{dI(\omega)}{d\Omega} = \frac{\omega^2}{16\pi^3}\left|\bd{\hat{r}} \times \int\displaylimits_{0}^{\infty} d t\, \bd{\beta}(t) \textrm{exp}[-i(\bd{k}\cdot \bd{r}_0(t) - \omega t)]\right|^2.\ee
The integral is zero from $-\infty$ to $0$ but non-zero from $0$ to $+\infty$.  The non-zero contribution comes because  $\bd{\beta}(t) = \bd{\beta}$ for $t>0$ with trajectory function $\bd{r}_0(t) = \bd{\beta}t$. Using $\bd{k} = \omega \bd{\hat{r}}$, 
\be \frac{dI(\omega)}{d\Omega} = \frac{\omega^2}{16\pi^3}\left|\bd{\hat{r}} \times \bd{\beta}\right|^2\left|\int\displaylimits_{0}^{\infty} d t\, \textrm{exp}[-i\omega(\bd{\hat{r}}\cdot \bd{\beta}-1)t)]\right|^2.\label{omega2}\ee
Notice the pre-factor $\omega^2$ frequency dependence which will ultimately cancel out after appropriate integration. The integral diverges at late times (upper limit), so we use a convergence regulator $e^{-\epsilon t}$ and set $\epsilon \rightarrow 0$ after integration. Using $\bd{\hat{r}}\cdot \bd{\beta} = \cos\theta$, where $\theta$ is the angle between $\bd{\beta}$ and the observation point, the integral is:
\be\int\displaylimits_{0}^{\infty} d t\, \textrm{exp}[-i\omega(\cos\theta-1)t-\epsilon t)] =  \frac{i}{i \epsilon+\omega(  1-\beta \cos \theta) },\ee
where we can now set $\epsilon \rightarrow 0$, and write the square of the integral as,
\be \left|\int\displaylimits_{0}^{\infty} d t\, \textrm{exp}[-i\omega(\bd{\hat{r}}\cdot \bd{\beta}-1)t)]\right|^2 =  \frac{1}{\omega^2(1- \beta \cos \theta )^2},\ee
demonstrating that frequency dependence cancels out exactly from Eq.~(\ref{omega2}).  Using $\bd{\hat{r}}\times \bd{\beta} = \beta \sin\theta$, the result is
\be \frac{dI(\omega)}{d\Omega} = \frac{1}{16\pi^3} \frac{\beta^2 \sin^2\theta}{(1-\beta\cos\theta)^2}.\ee
This is angular distribution of energy radiated per unit frequency in the frequency domain for the abrupt creation of a moving mirror in $3+1$-dimensions.  Applied to the simple case of violent acceleration, we see that the angular distribution of energy radiated per unit frequency, (and the total energy radiated per unit frequency in the next subsection) are independent of frequency.  
\subsubsection*{Total energy \& particles}
\begin{figure}[ht]
\begin{center}
{\rotatebox{0}{\includegraphics[width=3.5in]{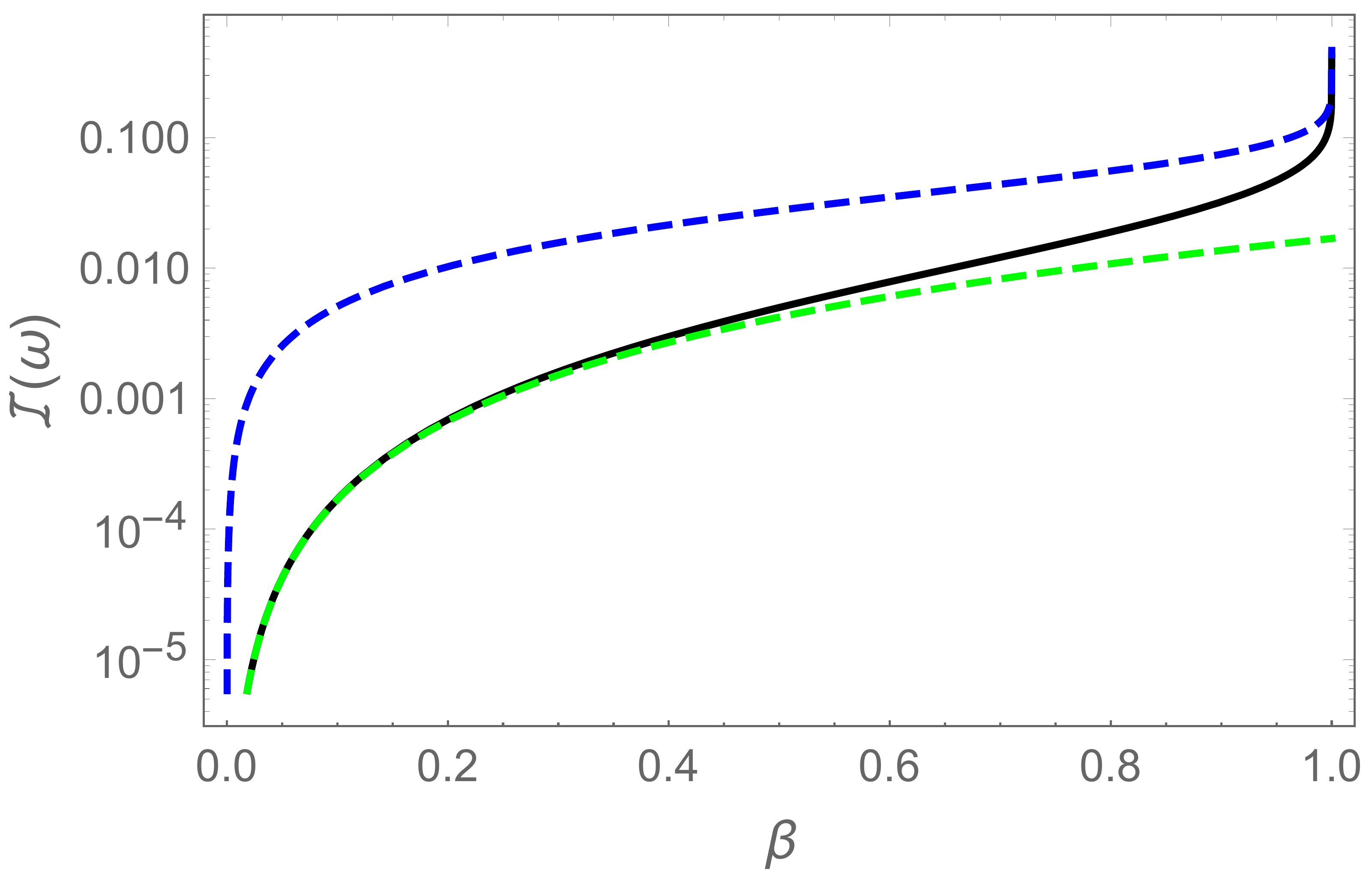}} {\caption{\label{fig2} A plot of the spectrum, $I(\omega)$, for the violent acceleration of a moving mirror to constant velocity, Eq.~(\ref{I(beta)}), as a function of the final speed, $\beta$.  Notice the radiated intensity is negligible at non-relativistic speeds (green) scaling as $\beta^2$ but at ultra-relativistic speeds (blue) the spectrum scales as the rapidity $\eta$. Eq.~(\ref{I(beta)}) corresponds to the spectrum of an electron's radiation during beta decay.    }}} 
\end{center}
\end{figure}

The total energy radiated per unit frequency is found by the integral
\be I(\omega) = \frac{1}{16\pi^3} \int d\Omega \frac{\beta^2 \sin^2\theta}{(1-\beta\cos\theta)^2}\ee
Using $d\Omega = \sin\theta d\theta\d\phi$, the numerator scales by $\sin^3\theta$ and integrating $\theta$ from $(0,\pi)$ and $\phi$ from $(0,2\pi)$, we obtain
a simple expression in terms of the final rapidity, $\eta =\tanh^{-1}\beta$, and final velocity, $\beta$, of the mirror,
\be I(\omega) = \frac{1}{2\pi^2}\left(\frac{\eta}{\beta} - 1\right).\label{I(beta)}\ee
For non-relativistic speeds $\beta \ll 1$, the radiated intensity is negligible and scales as 
$I(\omega) =\beta ^2/(6 \pi^2)$.  For ultra-relativistic speeds, $I(\omega) = \eta/(2\pi^2)$.  See Figure \ref{fig2} for a plot of Eq.~(\ref{I(beta)}).  
Ultimately, the spectrum does not depend on the frequency $\omega$ because the mirror is made to be in instantaneous motion at $t=0$ with velocity $\bd{v}$.  A more physical picture will have the velocity approached in some very short time interval $\Delta t$.  In this case the spectrum will die off and be negligible when $\omega \gg 1/\Delta t$.
The number of scalars per unit energy range is given by
\be N(\omega) = \frac{1}{2\pi^2\omega}\left(\frac{\eta}{\beta}-1\right).\ee
Their total energy radiated has a maximum frequency dependence:
\be E_\textrm{rad} = \int\displaylimits_0^{\omega_{\textrm{max}}} \omega N(\omega) d\omega = \frac{1}{2\pi^2}\left(\frac{\eta}{\beta}-1\right)\omega_{\textrm{max}}.\ee
We reiterate that the mirror is assumed to be created with constant velocity such that the acceleration period occurs in a very short time interval.  The spectrum function is negligibly small when the frequency is much larger than the inverse time interval of acceleration.  The intensity distribution, Eq.~(\ref{I(beta)}), is the same form as typical inner bremsstrahlung  spectrum \cite{Jackson:490457} in classical electrodynamics. In the case of beta decay \cite{Jackson:490457,Zangwill:1507229} e.g., the electron plays the role of the mirror.   
\section{Discussion}\label{sec:dis}
First, we should comment on the regime of validity of $P = \alpha^2/(6\pi)$ and the assumptions that were required to obtain it.  Eq.~(\ref{parts1}) has undergone an integration by parts, (for details see Eq. 2.34 of \cite{good2013time}), which is only valid if globally, the acceleration of the mirror is asymptotically zero in both the past and future.  Moreover, the boundary terms only disappear if the mirror is sub-light speed asymptotically, which is an even stronger constraint.  

This seemingly inconsequential subtlety almost certainly has much to do with the long standing debate \cite{Thomas1960,David1980} over whether a uniformly accelerated point charge radiates \footnote{Consider the clear distinctions needed  between the electromagnetic power received by a set of far-off observers and the instantaneous mechanical power loss of the charge in \cite{singal2020discrepancy}.}; in this case we comment that the power formula does not apply to global uniform acceleration as that would violate the previously mentioned assumptions giving non-vanishing boundary terms. That is, a globally uniformly accelerated mirror does not radiate \cite{Birrell:1982ix} energy but does radiate particles (as is well known \cite{Kay:2015iwa,Obadia:2002ch,Obadia:2002qe}). This is also in-line with the fact that uniformly accelerated point-like structure-less sources emit only zero-energy Rindler particles \cite{1819893}.

Second, we comment on the assumption that the Lorentz power scalar holds in higher dimensions.  Conformal invariance breaks down in higher dimensions \cite{Chen:2020sir} but our derivation, underscoring the Larmor formula as a Lorentz scalar independent of direction, does not ostensibly require conformal invariance.  The moving mirror in 1+1 dimensions permits exact solutions to the field equation because of conformal invariance but our result does not appeal to an exact explicit solution to the wave equation of motion.  It only appeals to the dynamics of the mirror as computed through the general renormalized quantum stress tensor as given by the Schwarzian derivative.  In general contexts, the quantum stress tensor, as opposed to particle production, is much easier to obtain.  This is true in higher dimensions where conformal invariance no longer holds (see \cite{Chen:2020sir} and references therein, e.g. Refs. 15-19).  

Third, we comment on the general applicability of the power distribution of Eq.~(\ref{dpdO}).  This formula applies to very general trajectories (albeit globally asymptotically sub-light speed  and time-like\footnote{Although this does not stop one from being able to locally compute the power distributions of the five classes of uniform acceleration, which for an electric charge will be the same form as for uniformly accelerated moving mirrors \cite{Good:2019aqd}.  It is unclear whether the effective temperatures will also carry-over \cite{Good:2020hav}.}).  For instance, a mirror moving on a circular arc near the speed of light will emit a synchrotron-type of radiation in the form of a narrow and intense beam directed tangent to the arc, implying that a fixed observer will see a brief flash or pulse of radiation every time the mirror moves directly toward them.  

This power distribution, Eq.~(\ref{dpdO}), in the context of vacuum acceleration radiation, warrants study because of its importance in connection to synchrotron radiation. Disentanglement in contexts where it is relevant will be essential in confirming the source's origin.  While astrophysical synchrotron radiation is a powerful indicator of the presence of magnetic fields and particle acceleration mechanisms near pulsars and black holes, the possibility that similar radiation could point to quantum amplification of vacuum fluctuations due to an accelerated boundary condition is of interest to a wide range of physicists concerned with relativistic quantum fields and information.   

\section{Conclusions}\label{sec:conc}
We have investigated the acceleration radiation emitted by a single relativistic perfect point mirror in $3+1$ dimensions and its analogies with a moving point charge in electrodynamics.  Namely,
\begin{itemize}
    \item We found the quantum power formula for moving mirrors and identified it with Larmor's form.
    \item The quantum Larmor formula is not applicable for eternal uniform acceleration. \item We generalized the $1+1$ dimensional moving mirror model to $3+1$-dimensions in the context of distributed power. This was done with the ansatz that the scalar power in $1+1$ dimensions (which is proportional to the proper acceleration magnitude squared) is also the scalar power in $3+1$ dimensions, i.e. the scalar will be invariant under Lorentz transformations in $3+1$ dimensions and obeys Larmor's form.  The only covariant choice of definition for scalar magnitude of proper acceleration in $3+1$ dimensions is that constructed by the usual Lorentz 4-vector acceleration. 
    
\item Consequently, we derived the power distribution and found its form is in analog to the well-known power distribution in electrodynamics (i.e. the apportioning responsible for synchrotron radiation, for example.).  This was done by Lorentz invariant vector decomposition of relativistic acceleration. The allotment is a result of the motion of the mirror only and the derivation did not rely on the use of fields or potentials i.e. no requisite Lienard-Wiechert potentials or electric and magnetic field counterparts for the quantum scalar radiation.
\item We derived the spectrum of a moving mirror abruptly created and violently accelerated to a constant velocity.  In analog to beta decay in classical electrodynamics, we found the appearance of the moving mirror plays the role of the moving electron.

\end{itemize}

 Knowing the radiated distribution of power for the moving mirror is a necessary first step toward precise orientation for accurate detection. The work here determines that the distribution for quantum scalar radiation from a moving mirror is the same form as that of accelerated electron radiation in classical electrodynamics.  The close analogy suggests direction for future work including application and interpretation of moving mirror trajectories in analog to the exactly known spectra  associated with the moving point charge including, for example, and not limited to, the well-known
bending magnet trajectory, the undulator trajectory, and the collinear acceleration burst trajectory.
\\
\\

\acknowledgments 

Funding from state-targeted program ``Center of Excellence for Fundamental and Applied Physics" (BR05236454) by the Ministry of Education and Science of the Republic of Kazakhstan is acknowledged, as well as the FY2018-SGP-1-STMM Faculty Development Competitive Research Grant No. 090118FD5350 at Nazarbayev University. 

\appendix
\section{Comment on Invariant Scaling}\label{LIscalar}

A Lorentz scalar is a number which is invariant under Lorentz transformations.  The most well-known examples include the speed of light $c$, the spacetime distance between two fixed events $\Delta s^2$, rest mass $m_0$, proper time $\tau$, and, $\bd{E}\cdot \bd{B}$ and $E^2-B^2$ in electrodynamics.

Numbers which are not invariant under Lorentz transformations, which we could call `non-Lorentz' scalars, may have no associated direction but nevertheless change under a Lorentz transformation.  Examples include $\bd{E}\cdot\bd{E}$ or electric charge density which are invariant with respect to spatial rotations but not with respect to boosts. Other examples are components of vectors and tensors which in general are altered under Lorentz transformations. 

A Lorentz scalar is invariant in a given dimensionality, but it is not a priori guaranteed that the same physics in a different dimension would render the same scaling.  Consider the Stefan-Boltzmann law which scales as $P \sim T^2$ in (1+1)D and $P\sim T^4$ in (3+1)D.  Despite no associated direction, the physics changes dramatically in different dimensions (see below for a caveat).  This may be an example of a number which is not invariant under Lorentz transformation - a `non-Lorentz' scalar (see e.g. a relativistic Stefan-Boltzmann law \cite{relSB} ).  There is no consensus on whether temperature is a Lorentz scalar.\footnote{Einstein \cite{ein} and Planck \cite{planck} derived a moving body to be cooler, $T' = T/\gamma$, while Ott \cite{ott} derived  a moving body to be warmer, $T' = \gamma T$.  Landsburg \cite{landsburg} derived $T' = T$.}

Our ansatz that the quantum power remains invariant under a change of dimensions, used to derive Eq.~(\ref{ansatz}), is motivated by the invariance of the proper acceleration as a Lorentz scalar.  This assumption is akin to conjecturing that the speed of light remains the same in both (1+1)D and (3+1)D. While our conjecture is no way guaranteed, it is a good starting point for the form of the (3+1)D quantum power.

The caveat to the Stefan-Boltzmann scaling is seen by considering a (1+1)D thermal system (like the thermal moving mirror \cite{carlitz1987reflections} with energy flux $\mathcal{F} = \pi T^2/12$), where the power is thus 
  \be P \sim T^2. \label{1Dpower} \ee
  However for a black body surface in (3+1)D, using the usual Stefan-Boltzmann law, we have 
  \be P \sim A T^4, \label{SB} \ee
  which, like we have just mentioned, seemingly deviates from Eq.~(\ref{1Dpower}).  However, for a Schwarzschild black hole, temperature $T \sim M^{-1}$, we see the area is $A\sim M^2$ or $A \sim T^{-2}$.  Substituting this into Eq.~(\ref{SB}) we retrieve Eq.~(\ref{1Dpower}). Therefore, a (3+1)D black hole acts\footnote{Accelerated boundary correspondences exist for more than just the Schwarzschild black hole \cite{Good:2016oey}. See the Reissner-Nordstr\"om \cite{good2020particle}, Taub-NUT \cite{Foo:2020xmy} and Kerr \cite{Good:2020fjz} black holes. De Sitter and anti-de Sitter cosmologies \cite{Good:2020byh} are also modeled by moving mirrors as well as extremal black holes \cite{Rothman:2000mm,good2020extreme,Liberati:2000sq}} as a (1+1)D moving mirror with the same power dependence on temperature.  This interesting caveat to the change in dimensional scaling provides an example where the power does remain invariant for a moving mirror.   
\section{Derivation of Perfect Differential Identity Eq.~(\ref{totald}) }\label{appA}
Starting with the right hand side of Eq.~(\ref{totald}),  where we have dropped the subscript of retarded time,
\begin{multline}
     \frac{d}{dt}\big[\frac{\bd{\hat{n}}\times(\bd{\hat{n}}\times\bd{\beta})}{1-\bd{\beta} \bd{\hat{n}}}\big]
    \\
    =\frac{(\bd{\hat{n}}\times(\bd{\hat{n}}\times\bd{\beta}))'(1-\bd{\beta} \bd{\hat{n}})-(1-\bd{\beta} \bd{\hat{n}})'(\bd{\hat{n}}\times(\bd{\hat{n}}\times\bd{\beta}))}{(1-\bd{\beta} \bd{\hat{n}})^2},
\end{multline}
using vector multiplication as the dot product (i.e. $\bd{\beta} \bd{\hat{n}} \equiv \bd{\beta} \cdot \bd{\hat{n}}$)  unless otherwise indicated.  Calculating the numerator using the bac-cab rule and then differentiating,
\begin{equation}
    [(\bd{\hat{n}}\times(\bd{\hat{n}}\times\bd{\beta})]'=[(\bd{\hat{n}}\bd{\beta})\bd{\hat{n}}-(\bd{\hat{n}} \bd{\hat{n}})\bd{\beta}]'=(\bd{\hat{n}}\dot{\bd{\beta}})\bd{\hat{n}}-\dot{\bd{\beta}},
\end{equation}
and differentiating the Doppler factor $g$,
\begin{equation}
    [1-\bd{\beta} \bd{\hat{n}}]'=-\dot{\bd{\beta}}\bd{\hat{n}},
\end{equation}
hence the numerator expanded out and simplified gives,
\begin{multline}
     ((\bd{\hat{n}}\dot{\bd{\beta}})\bd{\hat{n}}-\dot{\bd{\beta}})(1-\bd{\beta} \bd{\hat{n}})-(-\dot{\bd{\beta}}\bd{\hat{n}})((\bd{\hat{n}}\bd{\beta})\bd{\hat{n}}-(\bd{\hat{n}} \bd{\hat{n}})\bd{\beta})
    \\
    =(\bd{\hat{n}}\dot{\bd{\beta}})\bd{\hat{n}}-(\bd{\beta} \bd{\hat{n}})(\dot{\bd{\beta}} \bd{\hat{n}})\bd{\hat{n}}-\dot{\bd{\beta}}+\dot{\bd{\beta}}(\bd{\beta} \bd{\hat{n}})+(\bd{\beta} \bd{\hat{n}})(\dot{\bd{\beta}} \bd{\hat{n}})\bd{\hat{n}}-(\dot{\bd{\beta}} \bd{\hat{n}} )\bd{\beta}
    \\
    =-\dot{\bd{\beta}}-(\dot{\bd{\beta}}\bd{\hat{n}})\bd{\beta}+(\bd{\hat{n}}\dot{\bd{\beta}})\bd{\hat{n}}+\dot{\bd{\beta}}(\bd{\beta} \bd{\hat{n}}),\label{num_rhs}
\end{multline}
such that our time derivative gives a form inversely proportional to the Doppler factor squared,
\begin{equation}
    \frac{d}{dt}\big[\frac{\bd{\hat{n}}\times(\bd{\hat{n}}\times\bd{\beta})}{1-\bd{\beta} \bd{\hat{n}}}\big]=\frac{-\dot{\bd{\beta}}-(\dot{\bd{\beta}}\bd{\hat{n}})\bd{\beta}+(\bd{\hat{n}}\dot{\bd{\beta}})\bd{\hat{n}}+\dot{\bd{\beta}}(\bd{\beta} \bd{\hat{n}})}{(1-\bd{\beta} \bd{\hat{n}})^2}.
\end{equation}

The denominators of both hand sides of Eq.~(\ref{totald}) are now equal. Consider the numerator only of the left hand side of Eq.~(\ref{totald}) and expand it,

\begin{multline}
      \bd{\hat{n}}\times[(\bd{\hat{n}}-\bd{\beta})\times\dot{\bd{\beta}}] =(\bd{\hat{n}}\dot{\bd{\beta}})(\bd{\hat{n}}-\bd{\beta})-(\bd{\hat{n}}(\bd{\hat{n}}-\bd{\beta}))\dot{\bd{\beta}}
      \\
      =(\bd{\hat{n}}\dot{\bd{\beta}})\bd{\hat{n}}-\dot{\bd{\beta}}(\bd{\hat{n}}\dot{\bd{\beta}})-[\dot{\bd{\beta}}-(\bd{\hat{n}}\bd{\beta})\dot{\bd{\beta}}]
      \\
      =-\dot{\bd{\beta}}-\bd{\beta}(\bd{\hat{n}}\dot{\bd{\beta}})+(\bd{\hat{n}}\bd{\beta})\dot{\bd{\beta}}+(\bd{\hat{n}}\dot{\bd{\beta}})\bd{\hat{n}}\label{num_lhs}
\end{multline}
One can see that Eq.~(\ref{num_rhs}) and Eq.~(\ref{num_lhs}) are equal, thus we obtain the initial identity Eq.~(\ref{totald}).

\section{Derivation of $\Theta^*(\theta,\phi,\chi)$ }\label{appB}
The velocity vector $\bd{\beta}$ is directed along $z^*$-axis and the acceleration vector   $\dot{\bd{\beta}^*}$ is lying in the $x^*z^*$- plane. As we mentioned earlier, the angle between velocity and acceleration, as well as the azimuth, will be constant.In the instantaneous rest frame, the angles of $\theta^*$ and $\Theta^*$ change, therefore we express:
 \be \dot{\bd{\beta^*}}=\dot{\beta^*}(\sin{\chi^*} \hat{x}^*+\cos{\chi^*} \hat{z}^*),\ee
and $\bd{\beta}=\beta \hat{z}^*$.  Also, we know that the instant unit vector is:
\be \hat{\bd{n}}^* = \sin\theta^*\cos\phi^* \hat{x}^* + \sin\theta^*\sin\phi^* \hat{y}^* + \cos\theta^*\hat{z}^*.\ee
The dot product between each of the elements gives us:
\be \hat{\bd{n}}^*\cdot\bd{\beta}= \beta \cos{\theta^*},\ee
and likewise for the velocity with the instantaneous acceleration vector,
\be \bd{\beta} \cdot \dot{\bd{\beta^*}}= \beta \dot{\beta^*} \cos{\chi^*} \ee
and the unit instant vector with the instant acceleration vector
\be \hat{\bd{n}}^*\cdot \dot{\bd{\beta^*}}=\dot{\beta^*} \cos{\Theta^*}=\dot{\beta^*}(\cos{\phi^*}\sin{\theta^*}\sin{\chi^*}+\cos{\theta^*}\cos{\chi^*})\ee
We then rewrite angles of the lab frame in terms of angles of the instantaneous rest frame: 
\be \cos{\Theta^*}= (\cos{\phi^*}\sin{\theta^*}\sin{\chi^*}+\cos{\theta^*}\cos{\chi^*}). \ee
Converting from the instant rest frame to the lab frame gives
 \be \cos{\Theta^*}= \frac{\cos{\phi}\sin{\theta}\sin{\chi}}{\gamma g}+\frac{(\cos{\theta}-\beta)\cos{\chi}}{g} \ee
which is just,
 \be 1-\sin^2{\Theta^*}= \left(\frac{\cos{\phi}\sin{\theta}\sin{\chi}}{\gamma g}+\frac{(\cos{\theta}-\beta)\cos{\chi}}{g}\right)^2 \ee
Finally we have the general relationship for $\Theta^*(\theta,\phi,\chi)$:
\be \sin^2\Theta^*=1-\frac{(\cos{\phi}\sin{\theta}\sin{\chi}-\gamma(v-\cos{\theta})\cos{\chi})^2}{\gamma^2 g^2}.\label{B9} \ee 

Eq.~(\ref{B9}) explicitly demonstrates how the angle $\Theta^*$ in the instantaneous rest frame is related to the lab frame through angles $\phi$, $\theta$ and $\chi$.


\newpage
\onecolumngrid
\section{Limits of Power Distribution}\label{appC}
The angular distribution of instantaneous radiated power can be written in two forms, both in Eq.~(\ref{dPdO}) and in Eq.~(\ref{dpdO}), despite their different appearance they are equivalent (see also \cite{kirk}). As a consistency check, we demonstrate their limits in the most well-known cases and confirm they are also the same. Firstly, Eq.~(\ref{dpdO}) is expanded and rewritten in terms of lab frame angles $\theta, \phi$ and $\chi$. It is decomposed as follows:
\begin{multline}
\frac{dP}{d\Omega} =\frac{1}{16\pi^2} \frac{[\hat{\bd{n}} \times ((\hat{\bd{n}}-\bd{\beta})\times \dot{\bd{\beta}})]^2}{(1-\beta\cos\theta)^5}=
\frac{1}{16\pi^2} \frac{[( \hat{\bd{n}}-\bd{\beta})( \hat{\bd{n}}\cdot\dot{\bd{\beta}})-\dot{\bd{\beta}}(1-\bd{\beta}\cdot \hat{\bd{n}})]^2}{(1-\beta\cos\theta)^5}=\\
\frac{1}{16\pi^2} \frac{( \hat{\bd{n}}\cdot\dot{\bd{\beta}})^2 (1-2 \hat{\bd{n}}\cdot\dot{\bd{\beta}}+\bd{\beta}^2)-2( \hat{\bd{n}}\cdot\dot{\bd{\beta}}-\bd{\beta}\cdot\dot{\bd{\beta}})(1-\bd{\beta}\cdot \hat{\bd{n}})( \hat{\bd{n}}\cdot\dot{\bd{\beta}})+\dot{\bd{\beta}}^2 (1-\bd{\beta}\cdot \hat{\bd{n}})^2 }{(1-\beta\cos\theta)^5}=\\
\frac{1}{16\pi^2} \frac{( \hat{\bd{n}}\cdot\dot{\bd{\beta}})^2 (\bd{\beta}^2-1)+2( \hat{\bd{n}}\cdot\dot{\bd{\beta}})(\bd{\beta}\cdot\dot{\bd{\beta}})(1-\bd{\beta}\cdot \hat{\bd{n}})+\dot{\bd{\beta}}^2 (1-\bd{\beta}\cdot \hat{\bd{n}})^2 }{(1-\beta\cos\theta)^5}.\label{C1}
\end{multline}
The coordinate system is chosen so that vector $\bd{\beta}$ is directed along the $z$-axis, and the acceleration vector $\dot{\bd{\beta}}$ lies in the $x\;z$-plane:
\be \bd{\beta}=\beta\;\hat{z},\quad \dot{\bd{\beta}}=\dot{\beta}\;(\sin{\chi} \;\hat{x}+\cos{\chi}\; \hat{z}),\label{C2}\ee
and direction of the radiation  $\hat{\bd{n}}$ is given by:
\be \hat{\bd{n}} = \sin\theta\cos\phi\; \hat{x} + \sin\theta \sin\phi \;\hat{y} + \cos\theta\; \hat{z}.\label{C3}\ee
Thus, the dot product between the elements $\bd{n},\bd {\beta}$ and $\dot{\bd{\beta}} $ give:
\begin{equation}
    \hat{\bd{n}}\cdot\bd{\beta}= \beta \cos{\theta},  \quad \bd{\beta} \cdot \dot{\bd{\beta}}= \beta \dot{\beta} \cos{\chi}\quad \text{and} \quad \hat{\bd{n}}\cdot \dot{\bd{\beta}}=\dot{\beta}\cos{\Theta}= \dot{\beta}(\cos{\phi}\sin{\theta}\sin{\chi}+\cos{\theta}\cos{\chi}).\label{C4}
\end{equation}
Then by substituting equations Eq.~(\ref{C2}-\ref{C4}) into the equation of angular distribution in the expanded form Eq.~(\ref{C1}), Eq.~(\ref{dpdO}) is rewritten as follows:
\begin{equation}
\frac{dP}{d\Omega} =
\frac{1}{16\pi^2} \frac{ \dot{\beta}^2\cos^2{\Theta}\; (\beta^2-1)+2 \dot{\beta}^2 \beta \cos{\Theta} \cos{\chi}(1-\beta\cos\theta)+\dot{\beta}^2 (1-\beta\cos\theta)^2 }{(1-\beta\cos\theta)^5}.\label{C5}
\end{equation}
Now consider the limit of the power distribution  Eq.~(\ref{C5}) in the case of parallel and perpendicular positions of the acceleration and velocity vectors. For $ \chi=0$ (i.e. $a||v$), where $\cos {\Theta}=\cos{\theta}$:
\be \frac{dP_{||}}{d\Omega} =
\frac{1}{16\pi^2} \frac{ \dot{\beta}^2\cos^2{\theta}\; (\beta^2-1)+2 \dot{\beta}^2 \beta \cos{\theta}(1-\beta\cos\theta)+\dot{\beta}^2 (1-\beta\cos\theta)^2 }{(1-\beta\cos\theta)^5}=\frac{1}{16\pi^2} \frac{\dot{\beta}^2\sin^2{\theta}}{(1-\beta\cos\theta)^5},\label{C6}
\ee
and for $\chi=\pi/2$ (i.e. $a\perp v$), where $\cos{\Theta}=\cos{\phi}\sin{\theta}$:
\be
\frac{dP_\perp}{d\Omega} =
\frac{1}{16\pi^2} \frac{ \dot{\beta}^2\cos^2{\phi}\;\sin^2{\theta}\; (\beta^2-1)+\dot{\beta}^2 (1-\beta\cos\theta)^2 }{(1-\beta\cos\theta)^5}=
\frac{\dot{\beta}^2}{16\pi^2 (1-\beta\cos\theta)^3} \left(1-\frac{\cos^2{\phi}\;\sin^2{\theta}}{\gamma^2(1-\beta\cos\theta)^2}\right).\label{C7}
\ee
The power distribution limits by using the Eq.~(\ref{dPdO}) in terms of angles, is found by applying Eq.~(\ref{B9}):  
\be \frac{dP}{d\Omega} =\frac{1}{16\pi^2} \frac{\gamma^2[\dot{\bd{\beta}}^2-(\bd{\beta}\times \dot{\bd{\beta}})^2]}{(1-\beta\cos\theta)^3}\sin^2\Theta^*(\theta,\phi,\chi)=\frac{1}{16\pi^2} \frac{\gamma^2[\dot{\beta}^2-(\beta \dot{\beta} \sin{\chi})^2]}{(1-\beta\cos\theta)^3}\sin^2\Theta^*(\theta,\phi,\chi).\label{C8}\ee

\be \frac{dP}{d\Omega} = \frac{\dot{\beta}^2}{16\pi^2} \frac{\gamma^2[1-\beta^2 \sin^2{\chi}]}{(1-\beta\cos\theta)^3}\left(1-\frac{(\cos{\phi}\sin{\theta}\sin{\chi}-\gamma(\beta-\cos{\theta})\cos{\chi})^2}{\gamma^2 (1-\beta\cos\theta)^2}\right) .\label{C9}\ee
For  $\chi=0$ (i.e. $a||v$), Eq.~(\ref{C9}) transforms to:
\be \frac{dP_{||}}{d\Omega} = \frac{\dot{\beta}^2}{16\pi^2} \frac{\gamma^2}{(1-\beta\cos\theta)^3}\left(1-\frac{(\beta-\cos{\theta})^2}{ (1-\beta\cos\theta)^2}\right)=\frac{1}{16\pi^2} \frac{\dot{\beta}^2\sin^2{\theta}}{(1-\beta\cos\theta)^5} ,\label{C10}\ee
and for $\chi=\pi/2$ (i.e. $a\perp v$), Eq.~(\ref{C9}) converts to:
\be
\frac{dP_\perp}{d\Omega} = \frac{\dot{\beta}^2}{16\pi^2} \frac{\gamma^2[1-\beta^2 ]}{(1-\beta\cos\theta)^3}\left(1-\frac{(\cos{\phi}\sin{\theta})^2}{\gamma^2 (1-\beta\cos\theta)^2}\right)=\frac{\dot{\beta}^2}{16\pi^2 (1-\beta\cos\theta)^3} \left(1-\frac{\cos^2{\phi}\sin^2{\theta}}{\gamma^2 (1-\beta\cos\theta)^2}\right).\label{C11} \ee
Comparing equations Eq.~(\ref{C6}) and Eq.~(\ref{C10}), as well as equations Eq.~(\ref{C7}) and Eq.~(\ref{C11}) , it can be seen that the limits of Eq.~(\ref{dPdO}) and Eq.~(\ref{dpdO}) are the same, which gives the correct results for both rectilinear (braking) and circular (cyclotron) distributions.
\twocolumngrid
\bibliography{main} 
\end{document}